\documentclass{emulateapj}

\newcommand{\hhh}{H$_3^+$}

\begin{document}

\title{Interstellar Metastable Helium Absorption\\ as a Probe of the Cosmic-Ray Ionization Rate}

\author{Nick Indriolo\altaffilmark{1},
L. M. Hobbs\altaffilmark{2},
K. H. Hinkle\altaffilmark{3},
Benjamin J. McCall\altaffilmark{1,4}}

\altaffiltext{1}{Department of Astronomy, University of Illinois at
Urbana-Champaign, Urbana, IL 61801}
\altaffiltext{2}{University of Chicago, Yerkes Observatory, Williams Bay, WI 53191}
\altaffiltext{3}{National Optical Astronomy Observatories, Tucson, AZ 85726}
\altaffiltext{4}{Department of Chemistry, University of Illinois at
Urbana-Champaign, Urbana, IL 61801}

\begin{abstract}
The ionization rate of interstellar material by cosmic rays has been a major source of controversy, with different estimates varying by three orders of magnitude.  Observational constraints of this rate have all depended on analyzing the chemistry of various molecules that are produced following cosmic-ray ionization, and in many cases these analyses contain significant uncertainties. Even in the simplest case (H$_3^+$) the derived ionization rate depends on an (uncertain) estimate of the absorption path length.  In this paper we examine the feasibility of inferring the cosmic-ray ionization rate using the 10830 \AA\ absorption line of metastable helium.  Observations through the diffuse clouds toward HD~183143 are presented, but yield only an upper limit on the metastable helium column density.  A thorough investigation of He$^+$ chemistry reveals that only a small fraction of He$^+$ will recombine into the triplet state and populate the metastable level.  In addition, excitation to the triplet manifold of helium by secondary electrons must be accounted for as it is the dominant mechanism which produces He* in some environments.  Incorporating these various formation and destruction pathways, we derive new equations for the steady state abundance of metastable helium.  Using these equations in concert with our observations, we find $\zeta_{\rm He}<1.2\times10^{-15}$~s$^{-1}$, an upper limit about 5 times larger than the ionization rate previously inferred for this sight line using H$_3^+$.  While observations of interstellar He* are extremely difficult at present, and the background chemistry is not nearly as simple as previously thought, potential future observations of metastable helium would provide an independent check on the cosmic-ray ionization rate derived from \hhh\ in diffuse molecular clouds, and, perhaps more importantly, allow the first direct measurements of the ionization rate in diffuse atomic clouds.

\end{abstract}


\section{INTRODUCTION}

\subsection{Motivation}

Over the past several decades, the assumed value of the cosmic-ray ionization rate of interstellar hydrogen has fluctuated up and down.  Various theories and models have predicted ionization rates from $10^{-18}$~s$^{-1}$ to $10^{-15}$~s$^{-1}$ in the diffuse interstellar medium \citep[e.g.][]{spi68,van86,web98,lep04}.  On the other hand, observations of molecules such as HD and OH typically resulted in estimates of the ionization rate that were on the order of $10^{-17}$~s$^{-1}$ \citep{odo74,bla77,bla78,har78a,har78b,fed96}.  However, these estimates depend on gas phase abundances of O, OH, D, and HD, values which are often difficult to measure precisely.
%on the analysis of fairly complex chemical networks with some uncertain parameters.
More recently, observations of \hhh\ have again revised the cosmic-ray ionization rate upward to a few times $10^{-16}$~s$^{-1}$ \citep{mcc03,ind07}.  Deriving the ionization rate from \hhh\ requires only one uncertain parameter, the absorption path length.  While the higher ionization rates derived from \hhh\ are becoming generally accepted \citep{dalgarno}, it is desirable to search for new observables which can offer independent and less uncertain estimates of the ionization rate.

\begin{figure}
\epsscale{1.0}
\plotone{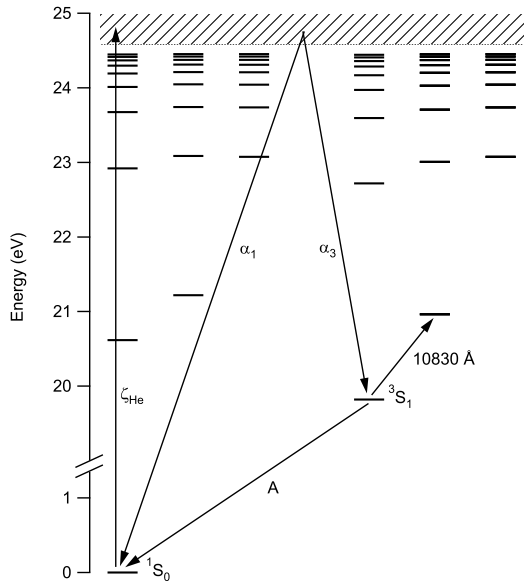}
\caption{Simplified energy level diagram of atomic helium, showing only S, P, and D terms up through a principal quantum number of 10.  The key processes thought to control the abundance of the metastable 2 $^3{\rm S}_1$ state, along with the absorption line from this state at 10830 \AA, are indicated.  It is assumed that all electron recombinations into either the singlet or triplet manifold quickly lead to the ground state of that manifold by allowed spontaneous emission.}
\label{EnergyLevels}
\end{figure}

In this report, we investigate the possibility of observationally determining the total ionization rate of helium atoms by cosmic rays in diffuse clouds.  The basic premise is that in a sufficiently reddened cloud, the column density of neutral helium atoms excited to the metastable 1s2s $^3$S$_1$ level may be high enough to be measured by means of interstellar absorption lines arising from that level. The high cosmic abundance of helium and the long radiative lifetime of the metastable level, $A^{-1}=2.5$ hr, may compensate for the difficulty of populating this highly excited level, which lies 19.8 eV above the ground level.
Previously, it has been assumed that this level should be populated almost entirely by cosmic-ray ionization of helium atoms, followed by radiative recombination of the ions with electrons.  Figure \ref{EnergyLevels} schematically shows the processes conventionally used in describing the (de)population of the metastable state.
%involved in populating the metastable state.

\subsection{Background}

A simple reaction network -- consisting of (i) cosmic-ray ionization of He$^0$ atoms in the 1s$^2$ $^1$S$_0$ ground level, (ii) radiative recombination of He$^+$ ions with electrons to the metastable 1s2s $^3$S$_1$ level, and (iii) forbidden spontaneous emission to the ground level -- was first used by \citet{sch68} and \citet{ree68} in proposing the observability of interstellar metastable helium.  However, both of these studies considered 2-photon emission from the metastable state, then thought to be the dominant method of depopulation.  This 2-photon decay proceeds on a timescale of $A^{-1}\gtrsim 116$~days \citep{dra68}.  In the following year, the radiative lifetime associated with 1-photon decay was computed to be $A^{-1}\approx 7$~hours \citep{gri69}, nearly 400 times faster than the 2-photon decay.  This value was later refined to $A^{-1}=2.5$~hours \citep{woodworth75,hata81}, still much faster, and as a result the analyses performed by \citet{sch68} and \citet{ree68} had greatly overestimated the population in the metastable state  \citep[this possibility was noted by][]{ree68}.  Consequently, the thought of observing interstellar metastable helium was abandoned.

Because of the high ionization rate inferred from \hhh, we decided to revisit these calculations considering up-to-date rate coefficients and improved telescope/detector capabilities.  Assuming the same chemical scheme as in the past, we can derive the steady state equations for the ground, ionized, and metastable states:
\begin{equation}
\alpha_{1}n_{e}n_{i}+n_{m}A=n_{g}\zeta_{\rm He},
\label{eqssg}
\end{equation}
\begin{equation}
n_{g}\zeta_{\rm He}=(\alpha_{1}+\alpha_{3})n_{e}n_{i},
\label{eqssi}
\end{equation}
\begin{equation}
\alpha_{3}n_{e}n_{i}=n_{m}A.
\label{eqssm}
\end{equation}
Here, $n_{g}$, $n_{i}$, and $n_m$ denote the populations of the ground, ionized, and metastable levels, respectively; $\zeta_{\rm He}$, the total ionization rate of He$^0$ atoms due to cosmic rays, including the effects of secondary electrons;  $\alpha_1$ and $\alpha_3$, the total, direct recombination rates to all singlet levels and to all triplet levels, respectively; $n_e$, the electron density in the gas; and $A$, the Einstein coefficient for spontaneous emission from the metastable level. The units of each term in equations (\ref{eqssg}) through (\ref{eqssm}) are cm$^{-3}$ s$^{-1}$. All recombinations to triplet levels above the metastable level are assumed to produce subsequent radiative cascades to the metastable level that are effectively instantaneous, owing to the long lifetime of the latter level. Similarly, all recombinations to singlet levels are assumed to cascade promptly to the ground level. Equations (\ref{eqssi}) and (\ref{eqssm}) can be solved for the ratios $n_i/n_g$ and $n_m/n_i$, and thus for $n_m/n_g$ as well. These ratios can be substituted into the definition of the fractional population of the metastable level,
\begin{equation}
f_m = \frac{n_m}{(n_m+n_i+n_g)},
\label{eqmetafrac}
\end{equation}
in order to obtain the desired relation between the fractional metastable population $f_m$ and the ionization rate $\zeta_{\rm He}$,
\begin{equation}
\frac{1}{f_m} = 1 + \frac{A}{(\alpha_3n_e)} + \frac{A}{(b\zeta_{\rm He})}.
\label{eqinvmetafrac}
\end{equation}
The triplet branching fraction for recombinations at 70~K is $b=\alpha_3/(\alpha_1+\alpha_3)=0.62$, since $\alpha_1=4.0\times10^{-12}$~cm$^3$~s$^{-1}$ and $\alpha_3=6.6\times10^{-12}$~cm$^3$~s$^{-1}$ (R.~Porter 2009, private communication).
Radiative decay to the ground level is by far the fastest of the three processes mentioned above, with $A=1.1\times10^{-4}$~s$^{-1}$. In contrast, $n_e=0.02$~cm$^{-3}$, $\alpha_3n_e=1.3\times10^{-13}$~s$^{-1}$, and $\zeta_{\rm He}=3\times10^{-16}$~s$^{-1}$ are representative values in diffuse clouds.  Given these values, $1\ll A/(\alpha_3n_e)\ll A/(b\zeta_{\rm He})$, and equation (\ref{eqinvmetafrac}) can be approximated by
\begin{equation}
\frac{1}{f_m}\approx\frac{A}{b\zeta_{\rm He}}.
\label{eqfmapprox}
\end{equation}
Owing to the very large differences among the rates, this approximation to equation (\ref{eqinvmetafrac}) is nearly exact.  This holds true as long as $\zeta_{\rm He}\ll n_e(\alpha_1+\alpha_3) \sim 10^{-13}$ s$^{-1}$, such that ionization of helium by cosmic-rays is the rate-limiting step in the path to the metastable state.  In this limit $f_m$
effectively depends on $\zeta_{\rm He}$ alone -- apart from the well-determined atomic constants $b$ and $A$ -- thus suggesting metastable helium as a fairly robust indicator of the cosmic-ray ionization rate.

\section{OBSERVATIONS}

\subsection{Predictions}

The fundamental question remaining then is whether the interstellar lines of He* arising from a suitable diffuse cloud are likely to be detectable. The strengths of these lines are fixed by the cloud's column density of metastable atoms, $N_m$, which can be calculated from
\begin{equation}
N_m = f_m N({\rm He}) = f_m A({\rm He}) N_{\rm H},
\label{eqmetacol}
\end{equation}
where $N({\rm He})$ is the total column density of helium atoms in all states, $N_{\rm H}$ is the total column density of hydrogen nuclei
[$N_{\rm H} = N({\rm H}) + 2N({\rm H}_2)$], and $A({\rm He})=N({\rm He})/N_{\rm H}=0.097$ is the relative abundance of helium with respect to hydrogen \citep{and89}.  The fraction of interstellar helium sequestered in the grains has also been assumed negligible. If a direct measurement of $N_{\rm H}$ is not available, an alternative is to use $N_{\rm H}=\beta E(B-V)$, where $E(B-V)$ is the observed color excess, and $\beta=N_{\rm H}/E(B-V)=5.8\times10^{21}$~cm$^{-2}$~mag$^{-1}$ is the interstellar gas-to-dust ratio \citep{boh78}.

To estimate the expected line strengths, we assume $E(B-V)=1.0$~mag and $\zeta_{\rm He}=3\times10^{-16}$~s$^{-1}$ in a suitable, individual interstellar cloud. The former value leads to $N_{\rm H}=5.8\times10^{21}$~cm$^{-2}$ and $N({\rm He})=5.6\times10^{20}$~cm$^{-2}$. A substitution of the assumed value of $\zeta_{\rm He}$ into equation (\ref{eqfmapprox}) gives $f_m=1.7\times10^{-12}$. Then, $N_m = f_m N({\rm He})=9.5\times10^{8}$~cm$^{-2}$.  The best choice among the available He~\textsc{i}* lines is the 1s2s $^3$S - 1s2p $^3$P multiplet located near 10830~\AA. Data for the transitions associated with this multiplet are shown in Table 1, where column 4 gives the oscillator strengths.  These lines are stronger than other transitions arising from the metastable level (such as the multiplet near 3889 \AA), and the near-infrared wavelength is advantageous in observations of heavily reddened stars with large total column densities of helium.

\begin{deluxetable}{cccc}
\tablecaption{The 1s2s $^3$S - 1s2p $^3$P Multiplet of He~\textsc{i} \label{tblmultiplet}}
\tablehead{\colhead{$\lambda_{air}$
} & & & \\
\colhead{(\AA)} & \colhead{$J$(lower)} & \colhead{$J$(upper)} & \colhead{$f$}
}
\startdata
10829.0911 & 1 & 0 & 0.060 \\
10830.2501 & 1 & 1 & 0.180 \\
10830.3398 & 1 & 2 & 0.300 \\
\enddata
\tablecomments{Wavelengths and oscillator strengths are from the NIST Atomic Spectra Database \citep{ral08}}
\end{deluxetable}

Assuming $N_m=9.5\times10^{8}$~cm$^{-2}$, the equivalent width of an unresolved blend of the two strongest lines of the multiplet, which are separated by only $2.5$~km~s$^{-1}$, would be $W_{\lambda}=0.47$~m\AA.
If a spectrometer with a resolving power of 70,000 were used, the line would have a central depth of $\sim$0.30\%, thus demanding a signal-to-noise ratio (S/N) of $\sim$1000 on the continuum for a 3$\sigma$ detection.  Modern optical echelle spectrographs can easily reach S/N exceeding 2000 \citep[e.g.,][]{abm03}, but reaching such a high S/N in the near-infrared is a significant challenge.

\subsection{Target Selection}

In choosing a target, we searched for sight lines that had a combination of several desirable characteristics: high color excess; high cosmic-ray ionization rate as inferred from H$_3^+$; relatively bright J-band magnitude; few interstellar velocity components; well-behaved stellar absorption features.  Using these criteria, we arrived at HD~183143 as our most favorable target, with $J=4.18$, $V=6.86$, $E(B-V)=1.27$, and a spectral type of B7Iae. The star's photospheric He~\textsc{i} absorption lines at 5875, 6678, and 7065~\AA\ are relatively broad, with ${\rm FWHM}\ge66$~km~s$^{-1}$, and they are also free from any interfering emission components \citep{tho03}. This suggests that the photospheric 10830~\AA\ line may desirably provide a smooth background with a shallow slope, against which one could search for the much narrower, weak interstellar lines. In addition, many interstellar absorption lines along this line of sight have been previously studied, including CH and CH$^+$ \citep{gre93}, and CN, \hhh, and $^{12}$CO \citep{mcc02}.  These observations reveal that there are two distinct interstellar cloud groups at different velocities.  Although this means that not all of the interstellar helium is at one velocity, it does provide a very useful method for potentially confirming a detection.  Additionally, \hhh\ observations have been used to determine the ionization rate of molecular hydrogen, $\zeta_2$, in this sight line \citep{ind07}.

Using values specific to the HD~183143 sight line ($E(B-V)=1.27; \zeta_{\rm He}=3.5\times10^{-16}$~s$^{-1}$) we can again perform the calculations in \S2.1 to determine the expected line strength.  The resulting equivalent width is $W_{\lambda}=0.70$~m\AA.  Because the sight line has 2 velocity components though, we assume equal amounts of material in each cloud, and thus expect 2 absorption lines with $W_{\lambda}=0.35$~m\AA.  These would require S/N$\sim$1300 for a 3$\sigma$ detection given the same instrument capabilities assumed above.  While obtaining a S/N this high is difficult in the near-infrared, some of the most advanced telescope/detector combinations are capable of approaching such results, so we proceeded with observations.

\subsection{Execution}

Data were taken near the He~\textsc{i}* line at 10830~\AA\ using the Phoenix spectrometer \citep{hin02} on the Gemini South Telescope.  The spectrometer was used with its echelle grating and 0.17'' slit to produce a resolving power of $\sim$70,000, and the J9232 filter to select the correct order.  Observations of both the target (HD 183143) and standard ($\alpha$ Aql) stars were taken on May 25, 2008 and June 28, 2008.  Total integration times were 33 and 30 minutes for the target and 1.9 and 1.4 minutes for the standard on each night, respectively.  During each set of observations, the star was nodded along the slit in an ABBA pattern to allow for the later subtraction of neighboring images, and thus the removal of the atmospheric background and detector bias levels.

\section{DATA REDUCTION}

A significant portion of the data reduction -- dark current subtraction, subtraction of neighboring images, removal of bad pixels, flat fielding, combination of exposures with the spectral image in the same nod position, fitting of the spectral response, and spectral extraction -- was performed using NOAO's IRAF package\footnote{See http://iraf.noao.edu/}.  During this process, we combined the methods outlined by \citet{kulth} with those given by NOAO's online Phoenix documentation\footnote{See http://www.noao.edu/usgp/phoenix/phoenix.html} in order to obtain the best possible S/N.  Once the one-dimensional spectra were extracted, they were imported to IGOR Pro\footnote{See http://www.wavemetrics.com/}, where we have macros set up to complete the reduction \citep{mccth}.

\begin{figure}
\epsscale{1.0}
\plotone{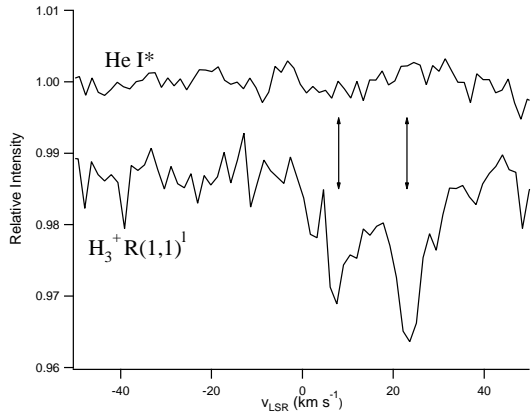}
\caption{Spectra of HD 183143 in velocity space.  The top spectrum, observed with Phoenix at Gemini South, has been adjusted for the centroid of the unresolved blend of the 1-1 and 1-2 members of the He~\textsc{i}* multiplet, and the broad photospheric line has been divided out.  The bottom spectrum \citep[from][]{mcc02} shows the $R(1,1)^l$ transition of H$_3^+$ for reference.  Arrows indicate the two interstellar velocity components which have been observed in various molecules (CH, CH$^+$, CN, $^{12}$CO, H$_3^+$).}
\label{figHespec}
\end{figure}

Because of the annual shift in the relative positions of (inter)stellar and atmospheric features with time, the data from different nights were first analyzed separately.  In all cases, however, the expected locations of the interstellar He~\textsc{i}* lines lie within the broad stellar absorption line.  Because the S/N of the standard star was actually {\it lower} than that of the target in the June data, we decided to forego the process of ratioing by the standard star and we instead directly fit the stellar absorption feature.  Both the A and B beams for each night were wavelength calibrated using atmospheric lines and then added together.  The broad stellar absorption feature was then fit using the summation of three gaussian functions, all of which were constrained to have FWHM at least 3 times that of the 10~km~s$^{-1}$ measured for interstellar absorption features along the line of sight.  The spectra from each night were then divided by their respective fits and shifted to be in the local standard of rest (LSR) frame.  Finally, the fully reduced spectra from both nights were added together and converted to velocity space to produce the top spectrum shown in Figure \ref{figHespec}.

\section{RESULTS}

There is no indication of interstellar He* absorption at either of the expected velocities.  While we did obtain a relatively high S/N ($\sim700$) for high-resolution infrared spectroscopy, we were unable to achieve the desired S/N $\sim1300$.  The non-detection of the He~\textsc{i}* lines enabled us to calculate an upper limit to the column density of metastable helium along this line of sight.  First, the upper limit to the equivalent width, $W_{\lambda}$, was computed via
\begin{equation}
W_{\lambda}<\sigma\lambda_{\rm pix}\sqrt{{\cal N}_{\rm pix}},
\label{eqeqwidth}
\end{equation}
where $\sigma=0.00145$ is the standard deviation in the spectrum, $\lambda_{\rm pix}=0.05$~\AA\ is the wavelength per pixel, and ${\cal N}_{\rm pix}=13$ is the number of pixels expected in a single absorption component given a 10~km~s$^{-1}$ FWHM (this is the average FWHM of \hhh\ lines reported by \citet{mcc02}).  These quantities result in an upper limit to the equivalent width of $W_{\lambda}<0.26$~m\AA\ ($W_{\lambda}<0.78$~m\AA\ at the 3$\sigma$ level).

Next, the standard relation between equivalent width and column density was used:
\begin{equation}
N=\frac{W_{\lambda}m_ec^2}{\lambda^2\pi e^2f},
\label{eqcolumn}
\end{equation}
where $m_e$ is the electron mass, $c$ is the speed of light, $\lambda$ is the wavelength of the transition, $e$ is the electron charge, and $f=0.48$ is the sum of the oscillator strengths of the 2 strongest, blended lines.  Inserting the known parameters results in a 3$\sigma$ upper limit to the column density $N_m<1.6\times10^9$~cm$^{-2}$ in a single velocity component.  However, there are 2 cloud components along this sight line, so the total line of sight limit is $N_m<3.2\times10^9$~cm$^{-2}$.

\section{ANALYSIS}

\subsection{Reaction Network Revisited}

In planning observations and calculating predicted line strengths, we relied on the simple chemical scheme that only considers the destruction of He$^+$ via electron recombination.  However, during the course of this study we identified (from advanced chemical models, e.g. \citet{woo07}) several competing reactions that could be important in destroying He$^+$ in diffuse clouds:
\begin{equation}
{\rm He^+} + {\rm H} \rightarrow {\rm He} + {\rm H^+},
\label{rHeH}
\end{equation}
\begin{equation}
{\rm He^+} + {\rm H_2} \rightarrow {\rm He} + {\rm H_2^+},
\label{rHeH2a}
\end{equation}
\begin{equation}
{\rm He^+} + {\rm H_2} \rightarrow {\rm He} + {\rm H} + {\rm H^+},
\label{rHeH2b}
\end{equation}
\begin{equation}
{\rm He^+} + {\rm CO} \rightarrow {\rm He} + {\rm O} + {\rm C^+}.
\label{rHeCO}
\end{equation}
Rate coefficients for these reactions, as well as the electron recombination reactions, can be determined for a specific temperature, $T$ (in Kelvin), using the fitting parameters ($\alpha$, $\beta$, and $\gamma$) in Table \ref{tblrates} in conjunction with the expression
\begin{equation}
k=\alpha\left(\frac{T}{300}\right)^{\beta}e^{-\gamma/T}~{\rm cm^3~s^{-1}}.
\label{eqalbega}
\end{equation}

Unlike the case of electron recombination, these reactions should not lead to metastable helium.  Simple energetics arguments demonstrate why this is the case.  The energy difference between the ionization potential of helium (24.6~eV) and the excitation energy of the metastable state (19.8~eV) is only 4.8~eV.  In order to dissociate and/or ionize the reaction partners of He$^+$, reactions (\ref{rHeH} -- \ref{rHeCO}) require energies of 13.6, 15.4, 18.1, and 22.5~eV, respectively (assuming all reactants are in the ground electronic state).  At diffuse cloud temperatures ($\sim$70~K) thermal energy is much less than 1~eV, and so the kinetic energy of the reactants will have no effect.  Following these arguments, the neutral helium product can only be in the ground state as more than 4.8~eV of the helium ionization potential must be used in each reaction.  As a result, accounting for these reactions greatly decreases the fraction of helium ions which pass through the triplet manifold, and thus the population of the metastable state.

However, we also made the assumption that the metastable state is only populated via cosmic-ray ionization of He, followed by electron recombination.  Given that a smaller branching fraction limits this pathway, electron impact excitation into the triplet manifold will be a competing formation mechanism.  Cross sections for ionization and excitation of helium by electrons in the $10-1000$~eV range are shown in \citet{dal99} figures 2a \& 2b.  To compute the rate of ionization and excitation, one must perform an integral in energy space over the product of each cross section with the differential energy spectrum of electrons in the ISM.  This full calculation is hindered by the fact that the spectrum of secondary electrons (those produced during ionization events) is unknown, and cannot be derived from the differential energy spectrum of cosmic-ray protons which is also unknown below $\sim1$~GeV.  The complexity associated with deriving the spectrum of secondary electrons is beyond the scope of this paper, thus we make some simplifications in estimating the importance of electron impact excitation into the triplet manifold of helium.

Assuming that all secondary electrons have the same energy, the ratio between the rate of excitation into all triplet states and the rate of ionization can be determined by taking the ratio of the respective cross sections at a given energy.  We take this ratio at 30~eV (the mean value given by \citet{cra78}), and find the rate of excitation into all triplet states to be 2 times faster than the rate of ionization by secondary electrons.  To determine the overall importance of electron impact excitation then, we need to find a relationship between the total ionization rate of helium and the ionization rate due to secondaries.  Using relations between the primary ionization rates of hydrogen and helium \citep{hab71,lis03} and between the primary ionization rate of hydrogen and the total ionization rate of helium \citep{gla74,tie05}, we estimate that ionization by secondary electrons accounts for about 1/6 of the total ionization rate of helium.  This, in turn, leads to the approximation that the rate for electron impact excitation into the triplet manifold --- and thus the metastable state (which we denote $\delta_{\rm He^*}$) --- should be roughly 1/3 that of the total ionization rate of helium (i.e. $\delta_{\rm He^*}\approx \zeta_{\rm He}/3$; we use this relation for the remainder of this paper).

%Assuming that all secondary electrons have energies of 30~eV (the average value given by \citet{cra78}), we can determine the ratio of the rate of excitation into all triplet states to the rate of ionization by comparing the respective cross sections at 30~eV.  In doing so, we find that the cross section for excitation is about 2 times that for ionization.  In order to relate the rate of excitation to the cosmic-ray ionization rate of helium due to protons, we use the relationship between the primary ionization rate of hydrogen and the total ionization rate of helium given by \citet{gla74}: $\zeta_{\rm He}=1.5\zeta_p$.
%If we assume that half of this difference in ionization rates is due to a difference between the ionization cross sections of H and He, then the other half is due to ionization by secondary electrons.  Because the ionization rate due to secondary electrons makes up 1/6 the total ionization rate, and the cross section for excitation into the triplet manifold is 2 times that of ionization, the rate for direct excitation to He$^*$ (which we denote $\delta_{\rm He^*}$) should be roughly one-third that of the total ionization rate of helium, i.e. $\delta_{\rm He^*}\approx \zeta_{\rm He}/3$.

%As the cross sections for ionization of hydrogen by protons \citep{bethe33} and %helium by protons \citep[and references therein]{pad09} are nearly equal

Mathematically, these additional formation and destruction reactions can easily be included by altering the steady state equations in \S1.2, resulting in 2 changes to our analysis.  First, due to the additional destruction pathways of He$^+$, the branching fraction, $b$, must be redefined as
\begin{equation}
b\equiv \frac{\alpha_{3}n_{e}}{n({\rm H})k_{\ref{rHeH}}\!+\!n({\rm H_2})(k_{\ref{rHeH2a}}\!+\!k_{\ref{rHeH2b}})\!+\!n({\rm CO})k_{\ref{rHeCO}}\!+\!n_{e}(\alpha_{1}\!+\!\alpha_{3})}.
\label{eqbranch}
\end{equation}
In many cases, however, absolute abundances are not known and it is thus convenient to recast equation (\ref{eqbranch}) in terms of fractional abundances as
\begin{equation}
b=\frac{\alpha_{3}x_{e}}{(1\!-\!f_{\rm H_2})k_{\ref{rHeH}}\!+\!f_{\rm H_2}(k_{\ref{rHeH2a}}\!+\!k_{\ref{rHeH2b}})/2\!+\!x_{\rm CO}k_{\ref{rHeCO}}\!+\!x_{e}(\alpha_{1}\!+\!\alpha_{3})},
\label{eqfbranch}
\end{equation}
where $x_j\equiv~n_j/n_{\rm H}$, $n_{\rm H}\equiv n({\rm H})+2n({\rm H_2})$, and the molecular hydrogen fraction $f_{\rm H_2}\equiv 2n({\rm H_2})/n_{\rm H}$.  Second, equation (\ref{eqfmapprox}) must be recast to include electron impact excitation into the metastable state, and becomes
\begin{equation}
\frac{1}{f_m}\approx\frac{A}{b\zeta_{\rm He}+\delta_{\rm He^*}}.
\label{eqfmBZDA}
\end{equation}

%where we have explicitly included the term $\delta_{\rm He^*}$, although we will assume $\delta_{\rm He^*}=\zeta_{\rm He}/3$ for the remainder of our analysis.

While the analysis now includes many more parameters, we can still calculate the fractional abundance of metastable helium, and thus the expected line strength, toward HD~183143. We assume that fractional abundances are constant throughout the cloud, allowing us to substitute column densities for number densities when available (i.e. $x_j=N_j/N_{\rm H}$). Using the color excess as in \S2.1 gives $N_{\rm H}=7.4\times10^{21}$~cm$^{-2}$.  This is used in conjunction with spectroscopic observations of CO which indicate $N({\rm CO})\approx 10^{15}$~cm$^{-2}$ \citep{mcc02} to compute $x_{\rm CO}$.  The assumption that there are equal amounts of atomic and molecular hydrogen is quantified by $f_{\rm H_2}=2/3$.  Finally, observations of C$^+$ in diffuse clouds  have shown that $x_e\sim1.4\times10^{-4}$, assuming that nearly all electrons come from this singly ionized carbon \citep{car96}. Combining these data and assumptions with the rate coefficients in Table \ref{tblrates}, the new branching fraction is $b=0.08$, about one-eighth of the value considering electrons alone.  Substituting this branching fraction and the relevant parameters from \S2.1-2.2 into equation (\ref{eqfmBZDA}) results in values of $f_m=1.3\times10^{-12}$, $N_m=9.3\times10^8$~cm$^{-2}$, and $W_{\lambda}=0.46$~m\AA.  Again splitting the material into 2 equal cloud components decreases the equivalent widths to $W_{\lambda}=0.23$~m\AA, which would require a S/N$\sim 2000$ for a 3$\sigma$ detection.

\subsection{Cosmic-Ray Ionization Rate of Helium}

Re-arranging equation (\ref{eqfmBZDA}),
%and assuming $\delta_{\rm He^*}=\zeta_{\rm He}/3$,
we can turn this problem around and compute an upper limit to the cosmic-ray ionization rate of helium using our observations.  Given the upper limit to the metastable column density, $N_m<3.2\times10^9$~cm$^{-2}$, and the estimated total helium column, $N({\rm He})=\beta A({\rm He})E(B-V)=7.1\times10^{20}$~cm$^{-2}$, the $3\sigma$ upper limit to the fractional metastable population is $f_m<4.5\times10^{-12}$.  Using this in concert with the branching fraction above, $b=0.08$, results in $\zeta_{\rm He}<1.2\times10^{-15}$~s$^{-1}$.  This upper limit is about 5 times larger than the ionization rate inferred from \hhh\ observations (assuming the relation between the ionization rate of helium and molecular hydrogen is given by $2.3\zeta_{\rm He}=1.5\zeta_2$ \citep{gla74}).  Because of electron impact excitation into the metastable state though, this determination of the ionization rate relies on a much more indirect analysis than was initially proposed.  Comparing $b\zeta_{\rm He}$ to $\delta_{\rm He^*}$, we can compute the relative importance of each formation mechanism via
\begin{equation}
P(\delta_{\rm He^*})=\frac{\delta_{\rm He^*}}{b\zeta_{\rm He}+\delta_{\rm He^*}}=(3b+1)^{-1}.
\label{eqdeltavszeta}
\end{equation}
In doing so, we find that electron impact excitation accounts for 80\% of the metastable population while ionization and electron recombination accounts for 20\%.

\section{DISCUSSION}

While the reactions associated with metastable helium are more complex than previously presented, we still see it as a viable tracer of the cosmic-ray ionization rate.  As such, we decided to investigate the prospects for He* detections in various interstellar environments, including diffuse molecular clouds ($100\lesssim n_{\rm H}\lesssim500$~cm$^{-3}$, $f_{\rm H_2}\gtrsim0.1$), dense clouds ($n_{\rm H}\gtrsim10^4$~cm$^{-3}$, $f_{\rm H_2}\approx1$), and diffuse atomic clouds ($n_{\rm H}\lesssim100$~cm$^{-3}$, $f_{\rm H_2}\lesssim0.1$) \citep{sno06}.  The following analyses will highlight the branching fraction in each environment, as well as the relative importance of electron impact excitation vs. ionization + recombination using equation (\ref{eqdeltavszeta}).

\subsection{Diffuse Molecular Clouds}

Given that the analysis in \S2.1 did not account for the processes examined in \S5.1, we felt it prudent to revisit the calculations for diffuse molecular clouds.  We use the same values as before ($E(B-V)=1$; $\zeta_{\rm He}=3\times10^{-16}$~s$^{-1}$), but now also assume $f_{\rm H_2}=2/3$, $x_e=1.4\times10^{-4}$, and $x_{\rm CO}=10^{-7}$.  The general results for this environment ($b=0.08$; $f_m=1.1\times10^{-12}$; $N_m=6.3\times10^8$~cm$^{-2}$; $W_{\lambda}=0.31$~m\AA) are similar to those for the specific diffuse molecular sight line HD~183143, with the differences due to the lower color excess.  Assuming that all of the material has the same velocity, metastable helium absorption should be observable in diffuse molecular clouds at a 3$\sigma$ level with S/N$\sim1500$.  Given the small branching fraction above, $P(\delta_{\rm He^*})=0.8$ and we conclude that metastable helium is predominantly formed via electron impact excitation in diffuse molecular clouds.

%While such a S/N is beyond the reach of current technology, future observations with next generation telescopes/detectors will provide an important measure of the cosmic-ray ionization rate independent of \hhh, and not subject to the uncertainty in the absorption path length.  In fact, it will be possible to combine the ionization rate derived from metastable helium with the observed column density of H$_3^+$ to directly infer the absorption path length and thereby the average cloud density.

\subsection{Dense Clouds}

Dense clouds, while providing a larger total helium column, have several characteristics detrimental to the formation of metastable helium.  The cosmic-ray ionization rate tends to be about 1 order of magnitude lower in dense clouds than in diffuse clouds \citep{dalgarno}.  Also, the fractional abundance of electrons is much lower, $x_e\approx4\times10^{-8}$, while the fractional abundance of CO is much higher, $x_{\rm CO}\approx1.4\times10^{-4}$ \citep{woo07}. Because $k_{\ref{rHeCO}}$ is so much larger than any of the other rate coefficients, collisions with CO will dominate the destruction of He$^+$ and equation (\ref{eqfbranch}) can be simplified to
\begin{equation}
b\approx\frac{x_e\alpha_3}{x_{\rm CO}k_{\rm CO}}.
\label{eqdenb}
\end{equation}
Given the fractional abundances above and the relevant rate coefficients ($\alpha_3$ and $k_{\rm CO}$ were computed for $T=40$~K), the branching fraction is $b\sim10^{-6}$.  As a result, $P(\delta_{\rm He^*})\approx1$, meaning that metastable helium is formed exclusively by electron impact excitation in dense clouds.  Even with this formation mechanism though, the expected equivalent width ($W_{\lambda}=0.13$~m\AA) and necessary S/N for a 3$\sigma$ detection (S/N$\sim3700$), coupled with the large attenuation of the background star's flux at 1~$\mu$m, make the detection of He* in dense clouds highly unlikely.

\subsection{Diffuse Atomic Clouds}

Diffuse atomic clouds, on the other hand, have negligible concentrations of H$_2$ and CO \citep{sno06} and presumably share the high ionization rate of diffuse molecular clouds.  In purely atomic conditions, electron recombination only has to compete with reaction (\ref{rHeH}) and equation (\ref{eqfbranch}) can be approximated as
\begin{equation}
b\approx\frac{x_e\alpha_3}{k_{\ref{rHeH}}+x_{e}(\alpha_{1}+\alpha_{3})}.
\label{atobranch}
\end{equation}
The simplified result for atomic clouds is then $b\approx0.40$, with a corresponding $P(\delta_{\rm He^*})=0.45$, meaning that ionization and electron impact excitation play roughly equal roles in forming metastable helium in such environments. Despite this branching fraction being closer to the ideal case of $b=0.62$, the low amount of material along such a sight line ($E(B-V)\sim0.1$) results in a predicted equivalent width of $W_{\lambda}\approx0.06$~m\AA.  However, there are some diffuse atomic sight lines with more favorable conditions.  One such candidate, $\sigma$~Sco, has $E(B-V)=0.40$ \citep{cla93} and thus a predicted equivalent width of $W_{\lambda}\approx0.22$~m\AA\ using equations (\ref{eqfmBZDA}) \& (\ref{atobranch}).  However, $\sigma$~Sco also has measured values of $N({\rm H})=2.4\times10^{21}$~cm$^{-2}$, $N({\rm H_2})=6.2\times10^{19}$~cm$^{-2}$ \citep{sav77}, and $N({\rm CO})=6.5\times10^{12}$~cm$^{-2}$ \citep{all90}, which correspond to $f_{\rm H_2}=0.049$ and $x_{\rm CO}=2.6\times10^{-9}$.  Using these values and equation (\ref{eqfbranch}), we can test the accuracy of equation (\ref{atobranch}) at small molecular fractions.  The result is $b=0.31$, or about a 30\% error in the approximation.  At $f_{\rm H_2}=0.15$, equation (\ref{atobranch}) overestimates $b$ by a factor of 2, so this approximation should only be applied for $f_{\rm H_2}\lesssim0.1$.  Taking the branching fraction from the full calculation, we predict an equivalent width of $W_{\lambda}\approx0.20$~m\AA, and a corresponding S/N $\sim2400$ necessary for a 3$\sigma$ detection.  If such a detection can be made, however, it will provide the exciting opportunity to probe the cosmic-ray ionization rate in an environment where \hhh\ observations cannot be made due to the low molecular fraction.

%For both the specific sight line toward $\delta$~Sco and diffuse atomic clouds in general, ionization and direct excitation play roughly equal roles in forming metastable helium.

%As in the case of diffuse molecular clouds, it seems the detection of He~\textsc{i}* in diffuse atomic clouds will require substantial improvements in instrument capabilities. However, such a detection will provide the exciting opportunity to directly probe the cosmic-ray ionization rate in an environment where \hhh\ observations cannot be made due to the low molecular fraction.

\section{CONCLUSIONS}

We have analyzed the possibility of detecting absorption lines due to interstellar metastable helium at 10830~\AA.  Observations toward the diffuse cloud sight line HD~183143 were taken, and a spectrum with S/N$\sim$700 was obtained, but no interstellar He~\textsc{i}* lines were detected.  In examining the chemistry associated with metastable helium, we have identified important formation and destruction pathways, and have derived new equations for the steady state analysis of He*.  While these reactions have been known for some time, this is the first instance where they have been applied to metastable helium chemistry.  Using our observations and the newly derived equations, we inferred an upper limit for the cosmic-ray ionization rate of helium which, although consistent with other studies, is about 5 times larger than previously inferred values.

To determine if future observations of interstellar He* are warranted, we predicted the S/N ratios necessary for 3$\sigma$ detections in various environments.  Diffuse molecular clouds are the most promising targets with S/N$\sim$1500 required, while favorable diffuse atomic clouds need S/N$\sim$2400.  While such observations are extremely challenging at present, advancements in telescope and near-infrared detector technology may one day make metastable helium a widely applicable probe of the cosmic-ray ionization rate.  In diffuse molecular clouds, He* will act as a cosmic-ray probe independent of \hhh, and together with \hhh\ it will also enable determination of the absorption path length and average cloud density.  He* observations will also be especially important for diffuse atomic clouds, where there are no other reliable tracers of the ionization rate.

\mbox{}

%************************************

The authors wish to thank Tom Kerr and the UKIRT Service Observing Programme for obtaining preliminary spectra near 10830 \AA\ using CGS4, Ryan Porter and Gary Ferland for supplying unpublished values of the recombination coefficients and for suggesting the possibility of electron impact excitation into the metastable state, and the anonymous referee for helpful suggestions and comments.  NI and BJM have been supported by NSF grant PHY 05-55486.

Based on observations obtained at the Gemini Observatory, which is operated by the Association of Universities for Research in Astronomy, Inc., under a cooperative agreement with the NSF on behalf of the Gemini partnership: the National Science Foundation (United States), the Science and Technology Facilities Council (United Kingdom), the National Research Council (Canada), CONICYT (Chile), the Australian Research Council (Australia), Minist\'{e}rio da Ci\^{e}ncia e Tecnologia (Brazil) and Ministerio de Ciencia, Tecnolog\'{i}a e Innovaci\'{o}n Productiva  (Argentina).
The Gemini/Phoenix spectra were obtained through program GS-2008A-Q-14.

The observations were obtained with the Phoenix infrared spectrograph, which was developed by the National Optical Astronomy Observatory.

\clearpage

\begin{deluxetable}{lllrcc}
\tablecaption{Rate Coefficients for Reactions Involving Ionized Helium
\label{tblrates}}
\tablehead{ & & & & \colhead{Coefficient at 70 K} & \\
\colhead{Reaction} & \colhead{$\alpha$} & \colhead{$\beta$} & \colhead{$\gamma$} & \colhead{(cm$^3$~s$^{-1}$)} & \colhead{References}
}
\startdata
He$^+$ + H $\rightarrow$ He + H$^+$ & $1.2\times10^{-15}$ & ~0.25 & 0 & $k_{\ref{rHeH}}=8.3\times10^{-16}$ & 1 \\
He$^+$ + H$_2$ $\rightarrow$ He + H$_2^+$ & $7.2\times10^{-15}$ & ~0 & 0 & $k_{\ref{rHeH2a}}=7.2\times10^{-15}$ & 2 \\
He$^+$ + H$_2$ $\rightarrow$ He + H + H$^+$ & $3.7\times10^{-15}$ & ~0 & 35 & $k_{\ref{rHeH2b}}=2.2\times10^{-14}$ & 2 \\
He$^+$ + CO $\rightarrow$ He + O + C$^+$ & $1.6\times10^{-9}$ & ~0 & 0 & $k_{\ref{rHeCO}}=1.6\times10^{-9}$ & 3,4 \\
He$^+$ + $e$ $\rightarrow$ He$(1~^1{\rm S})$ + $h\nu$ & $1.76\times10^{-12}$ & -0.56 & 0 & $\alpha_1=4.0\times10^{-12}$ & 5 \\
He$^+$ + $e$ $\rightarrow$ He$(2~^3{\rm S})$ + $h\nu$ & $2.84\times10^{-12}$ & -0.59 & 0 & $\alpha_3=6.6\times10^{-12}$ & 5 \\
\enddata
\tablecomments{Coefficients at temperatures between about 10 and 300~K can be derived using $\alpha$, $\beta$, $\gamma$, and equation (\ref{eqalbega}).  Rate coefficients and their references for reactions (\ref{rHeH}), (\ref{rHeH2a}), (\ref{rHeH2b}), and (\ref{rHeCO}) were found at http://www.udfa.net/.}
\tablerefs{(1) \citet{sta98}; (2) \citet{barth}; (3) \citet{lau74}; (4) \citet{ani77}; (5) R.~Porter 2009, private communication }
\end{deluxetable}

%%%%%%%%%%%%%%%%%%%%%%%%%%%%%%%%%%%%%%%%%%%%%%%%%%%%%%%%%%%%%%%%%%%%%%%%%%

\end{document}